\documentclass[12pt]{article}
\usepackage{amsmath,amssymb,amsthm,color}
\usepackage{graphicx}
\usepackage{cite}
\usepackage{epsfig}
\usepackage{float}

\renewcommand{\baselinestretch}{1.66}
\begin{document}

\title {Stacking-enriched magneto-transport properties of few-layer graphenes\\ }

\author{
\small Thi-Nga Do$^{a,\dag}$, Cheng-Peng Chang $^{b}$, Po-Hsin Shih$^{a}$, Ming-Fa Lin$^{a,*}$ $$\\
\small  $^a$ Department of Physics, National Cheng Kung University, Tainan, Taiwan 701\\
\small  $^b$ Center for General Education,Tainan University of Technology, Tainan, Taiwan 710\\
 }

\renewcommand{\baselinestretch}{1.66}
\maketitle

\renewcommand{\baselinestretch}{1.66}

\begin{abstract}

The quantum Hall effects in the sliding bilayer graphene and AAB-stacked trilayer  system are investigated by the Kubo formula and the generalized tight-binding model.
The various stacking configurations can greatly diversify the magnetic quantization and thus create the rich and unique transport properties.
The quantum conductivities are very sensitive to the Fermi energy and magnetic-field strength. The diverse features cover the specific non-integer conductivities, the integer conductivities with the distinct steps, the splitting-created reduction and complexity of quantum conductivity, a vanishing or non-zero conductivity at the neutral point, and the well-like, staircase, composite, and abnormal plateau structures in the field-dependencies. Such stacking-dependent characteristics mainly originate from the crossing, anticrossing and splitting Landau-level energy spectra and three kinds of quantized modes.

\vskip 1.0 truecm

\par\noindent \dag Corresponding author. {~ Tel:~ +886-6-275-7575;~ Fax:~+886-6-74-7995.}\\~{{\it E-mail address}: ngado@phys.ncku.edu.tw (T.N. Do)}
\par\noindent * Corresponding author. {~ Tel:~ +886-6-275-7575;~ Fax:~+886-6-74-7995.}\\~{{\it E-mail address}: mflin@mail.ncku.edu.tw (M.F. Lin)}

\end{abstract}

\pagebreak
\renewcommand{\baselinestretch}{2}
\newpage

\section{\bf Introduction}
\vskip 0.3 truecm

Since the discovery of graphene, its unconventional quantum Hall effect (QHE) has attracted a lot of theoretical and experimental studies. \cite{S306;666, N438;197, N438;201, PRB65;245420, PRL95;146801, PRB73;125411, PRL96;256602, PRB74;075422, PRB74;161407, PRB74;195429, PE40;269} Few-layer graphene in a specific stacking configuration can be synthesized from highly orientated pyrolytic graphite,\cite{C49;3242, APL96;201909, PRB48;17427, PRB75;235449} metalorganic chemical vapor deposition (CVD),\cite{SR3;03143, JCG312;3219, C48;3169, JAP110;013720, NRL6;95, NL12;5539} chemically and electrochemically reduced graphene oxides,\cite{PNAS107;14999, NP7;953, PRB79;125411} arc discharge,\cite{NR3;611, ACSN3;411} and growth on (111) surface of diamond.\cite{JCP129;234709, JAP109;093523} Recently, large-area graphenes with high mobility and highly symmetric configurations, e.g., AA, AB and ABC stackings, have been found in CVD-grown samples.\cite{JCP129;234709, JAP109;093523, PRL102;015501, PRB81;161410, ACSN2012} AAB stacking \cite{PRB86;085428, SS601;498, APL102;163111, PRB79;125411, APL107;263101, PRB75;235449} and an intermediate bilayer stacking under the interlayer shift/twist, are also identified and becoming an interesting subject.\cite{NM8;203, NL8;4320, PRL107;216602, PRB85;201408} Particularly, the former can be produced by the horizontal shift of the top graphite layer with scanning tunneling microscopy (STM),\cite{PRB86;085428} freshly cleaving the sample with either a scalper or scotch tape,\cite{SS601;498} the liquid phase exfoliation of natural graphite in N-methyl-2-pyrrolidone and dispersed onto highly-oriented pyrolytic graphite,\cite{APL102;163111} and the epitaxial growth on SiC \cite{PRB79;125411} and Ru(0001) \cite{APL107;263101} substrates. Moreover, the lower-symmetry graphene layers could be generated from the atmospheric pressure graphitization of silicon carbide,\cite{NM8;203, PRL107;216602} and the growth on SiC \cite{NL8;4320} and Cu \cite{PRB85;201408} substrates with solid state graphitization and CVD, respectively. Both AAB-stacked trilayer graphene and sliding bilayer systems are suitable in studying the configuration-enriched QHE.\\

On the theoretical side, the generalized tight-binding model is developed to study the essential properties of the sliding bilayer graphene and AAB-stacked trilayer system, especially for the magnetically quantized features.\cite{C94;619, PCCP18;17597, CR4;7509, PCCP17;26008} The relative shift of two graphene layers can create the various stacking configurations (e.g., those of the transformation between AA and AB stackings) and thus induce the diverse magnetic quantization phenomena.\cite{CR4;7509} Under a uniform perpendicular magnetic field (${B_0\hat z}$), the sliding bilayer systems possess two groups of valence and conduction Landau levels (LLs). Furthermore, they present three kinds of LLs, namely, the well-defined, perturbed and undefined ones. Such LLs are characterized by the quantized oscillation modes in the spatial distributions. Furthermore, the rich and unique field-strength dependences are revealed in LL energy spectra, such as, the initiated  energies of distinct LL groups, the monotonous or non-monotonous ${B_0}$-dependencies, and the crossing/anti-crossing behaviors. These are attributed to the stacking-dependent interlayer atomic interactions of carbon 2p$_z$ orbitals. As for the AAB-stacked trilayer graphene, the lower stacking symmetry leads to the unusual band structure and LLs, mainly owing to the very complicated interlayer atomic interactions.\cite{C94;619} Three pairs of valence and conduction bands cover the oscillatory, sombrero-shaped and parabolic energy dispersions, leading to three groups of unusual LLs. State degeneracy of each LL is reduced to  half in the absence of mirror  symmetry about $z=0$ plane. Specifically, the $B_0$-dependent LL spectra exhibit the frequent anti-crossings arising from the localized  wave functions with the main and side modes. The stacking-diversified magneto-electronic properties are expected to dominate the quantum Hall conductivities. \\

There are a lot of experimental and theoretical studies on the rich and unique QHE in layered graphenes, especially for monolayer, AB and ABC stacking systems. Magnetic transport measurements on monolayer graphene \cite{N438;197, N438;201} have identified the unconventional half-integer Hall conductivity $\sigma_{xy}=(m+1/2)4e^2/h$, in which $m$ is an integer and the factor of 4 stands for the spin and sublattice-dependent degeneracy. This unusual quantization is attributed to the quantum anomaly of the ${n=0}$ LL corresponding to the Dirac point.\cite{PRL95;146801} As for AB-stacked bilayer graphene, the Hall conductivity is confirmed to be $\sigma_{xy}=4m^{'} e^2/h$ ($m^{'}$ a non-zero integer).\cite{PRB78;033403, NP2;177} Furthermore, there exists an unusual integer quantum Hall conductivity, a double step of $\sigma_{xy}=8e^2/h$, at zero energy and low magnetic field.\cite{PRL96;086805} This mainly comes from ${n=0}$ and 1 LLs of the first group.
The low-energy QHE plateaus of ABA-stacked trilayer graphene are observed at $\pm 2e^2/h$, $\pm4e^2/h$, $\pm6e^2/h$, and $\pm8e^2/h$ with a step height of $2e^2/h$, especially for the energy range of $\sim \pm 20$ meV.\cite{N7;621, NP7;948}
This observation is consistent with the calculated LL energy spectra.\cite{PRB84;125455} The neighboring and next-neighboring interlayer atomic interactions, respectively, create  the separated Dirac cone and parabolic bands, and the valley splitting of the latter.
At low energy, this further leads to six quantized LLs with double spin degeneracy and thus the QHE step of 2$e^2/h$. However, the higher-energy QHE in ABA stacking could be regarded as the superposition of those of monolayer and AB bilayer.\cite{N7;621, NP7;948, PRB84;125455}
The ABC-stacked trilayer graphene, on the other hand, presents the important differences in the main features of QHE compared with the ABA trilayer system. The quantum Hall conductivity is quantized as a sequence of $\sigma_{xy}= 4(\pm |m^{'}| \pm 1/2)e^2/h$ in the absence of the $\sigma_{xy}=\pm 2e^2/h$, $\pm 4e^2/h$, and $\pm 8e^2/h$ plateaus.\cite{NP7;948, NP7;953}
Specifically, a $\sigma_{xy}=12e^2/h$ step appears near zero energy, being associated with the $n=0$, 1 and 2 LLs of the first group due to the surface-localized flat bands.\cite{PCCP17;15921}
In fact, the fourfold spin and valley degeneracy is retained for the ABC trilayer due to the inversion symmetry, resulting in the QHE step height of $4e^2/h$.
The above-mentioned interesting phenomena of electronic transport properties in graphene opens the door to exploring the configuration-enriched QHE in other layered graphenes, such as AAB-stacked trilayer and sliding bilayer systems.\\

In this work, the linear Kubo formula, combined with the generalized tight-binding model, is used to investigate the unusual QHE in few-layer graphenes with high- and low-symmetry stacking configurations.
The developed model would be very useful in the identification of the magneto-electronic selection rules under the static case; that is, the available transition channels in magneto-transport property could be examined thoroughly.
The dependencies of quantum conductivity on the Fermi energy ($E_F$) and magnetic-field strength are explored in detail.
This study shows that the feature-rich LLs can create the extraordinary magneto-transport properties.
The sliding bilayer systems, with three kinds of LLs present the unusual QHE, covering the integer and non-integer conductivities, the zero and non-vanishing conductivities at the neutral point, the well-like, staircase and composite  quantum structures, and the different step heights. Furthermore, the reduced conductivity,  the complex plateaus and the abnormal structures are revealed in the AAB-stacked trilayer graphene, with the splitting and perturbed LLs.
These results are deduced to be dominated by  the crossing and anti-crossing energy spectra, and the spatial oscillation modes.
The predictions on the diverse quantum conductivities could be experimentally verified, as done for other layered graphene.\cite{N438;197, N438;201, PRB78;033403, NP2;177, N7;621, NP7;948}\\

\vskip 0.6 truecm
\par\noindent
\section{\bf Models and Methods }
\vskip 0.3 truecm

The generalized tight-binding model is developed to investigate the essential properties of layered graphenes in external fields.
The low-energy Hamiltonian is mainly built from the 2$p_z$-orbital tight-binding functions in a unit cell. The two sublattices in the $l$-th $(l=1,2,...,N)$ layer are denoted as $A^{l}$ and $B^{l}$, where $N$ is the number of graphene layers.
The interlayer distance and the C-C bond length are, respectively, $d_0=3.37 \AA$ and $b_0=1.42$ $\AA$.
The various Hamiltonian matrices are dominated by the intralayer and the interlayer atomic interactions.
For sliding bilayer graphene, the stacking configurations can be transformed according to AA ($\delta=0$) ${\rightarrow}$ AB (${\delta\,=b_0}$) ${\rightarrow}$ AA$'$ ($\delta=1.5b_0$) by the relative shift between two graphene layers along the armchair direction ($\hat x$). The strength of the hopping integrals $\epsilon_{i,j}$ between the lattice sites $i$ and $j$ is dependent on the distance and the angle of two 2$p_z$ orbitals obeying the following relation \cite{PRB88;115409}:
\begin{align}
-\epsilon_{ij}= \epsilon_{0} e^{-\frac{d-b_0}{\varrho}}[1-(\frac{\mathbf{d}.\mathbf{e_z}}{d})^2]
+ \epsilon_{1} e^{-\frac{d-d_0}{\varrho}}(\frac{\mathbf{d}.\mathbf{e_z}}{d})^2.
\end {align}
$\epsilon_{0}=2.7$ eV and $\epsilon_{1}=0.48$ eV are, respectively, the intralayer nearest-neighbor and interlayer vertical hopping integrals, $\mathbf{d}$ the position vector connecting two lattice sites, and $\varrho=0.184b_0$ the decay length.\cite{CR4;7509}\\

For trilayer graphenes, AAB stacking is chosen for a model theoretical  study. In this system, the A atoms in three layers have the same (x,y) coordinates, while the B atoms on the third layer are projected at the hexagonal centers of the other two layers. There exist 10 kinds of atom-atom interactions in creating the unusual energy dispersions; that is, they are responsible for the oscillatory, sombrero-shaped and parabolic bands.\cite{C94;619} $\gamma_{0}=-2.569  $ eV represents the nearest-neighbor intralayer atomic interaction; $ \gamma_{1}=-0.263$ eV, $\gamma_{2}=0.32 $ eV, $ \gamma_{3}=-0.413$ eV, $\gamma_{4}=-0.177 $ eV, $ \gamma_{5}=-0.319$  eV, $ \gamma_{6}=-0.013  $ eV, $ \gamma_{7}=-0.0177 $ eV, and $ \gamma_{8}=-0.0319  $ eV stand for the interlayer atomic interactions among three layers. $ \gamma_{9}=-0.012 $ eV accounts for the difference in the chemical environment of A and B atoms. Such critical interactions can greatly complicate the magnetic Hamiltonian and thus diversify the magneto-electronic properties. In addition, the calculated band structures are almost identical to those evaluated from the first-principles calculations.\cite{C94;619}\\

\begin{figure}[H]
\center
\rotatebox{0} {\includegraphics[width=0.9\linewidth]{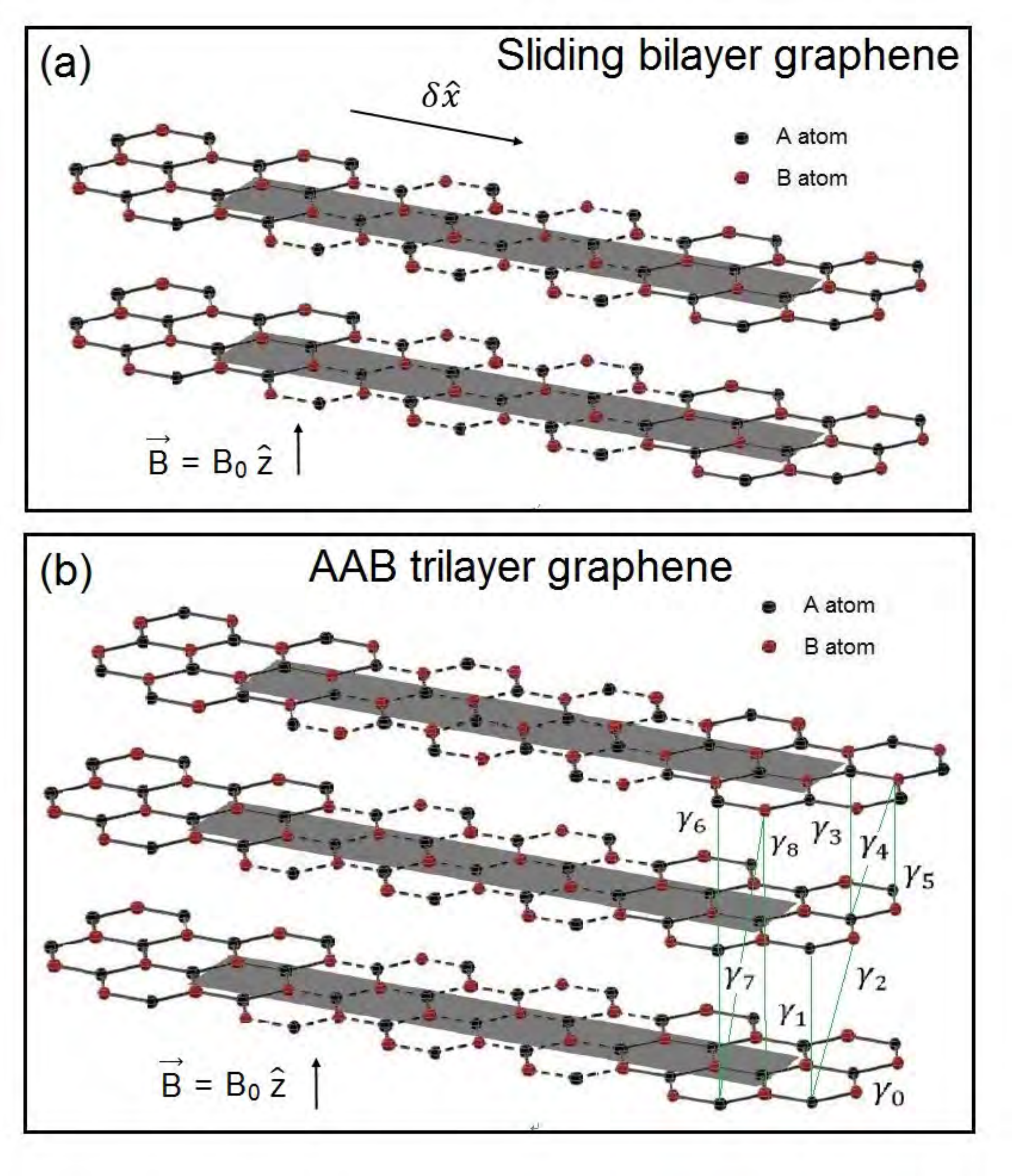}}
\caption{Geometric structure of (a) sliding bilayer graphene and (b) AAB trilayer graphene with an enlarged unit cell in the presence of a uniform perpendicular magnetic field.}
\label{Figure 1}
\end{figure}

When layered graphenes exist in a uniform perpendicular magnetic field, the quantized LLs could be explored thoroughly using the generalized tight-binding model even for low-symmetry stacking configurations. This model is based on the subenvelope functions of of the distinct sublattices, in which the magnetic Hamiltonian is built from the bases of tight-binding functions coupled with a Peierls phase factor. A zero-field hexagonal unit cell, with 2$N$ carbon atoms, is changed into an enlarged rectangular cell including 4${NR_B}$ carbon atoms (Fig. 1(a) for bilayer graphene), since the vector potential (${{\bf A}=[0,B_0x,0]}$) can induce a periodical Peierls phase characterized as the path integration of ${\bf A}$ between two lattice sites  (detailed calculations with various interlayer atomic interactions in Ref. 42). $R_B$ is the ratio of the flux quantum (${\phi_0\,=hc/e}$) versus the magnetic flux through a hexagon (${\phi\,=3\sqrt 3\,b_0^2B_0/2}$), e.g.,  ${R_B=8\times\,10^3}$ at $B_0$ = 40 T.\\

The magnetically quantized LLs in layer graphenes can create the unique transport properties. Within the linear response, the transverse Hall conductivity is evaluated from the linear Kubo formula.\cite{JAP112;044306}
\begin{align}
\sigma_{xy} = \frac {ie^2 \hbar} {S}
&\sum_{\alpha} \sum_{\alpha \neq \beta} (f_{\alpha} - f_{\beta})
\frac {\langle \alpha  |\mathbf{\dot{u}}_{x}| \beta\rangle  \langle \beta |\mathbf{\dot{u}}_{y}|\alpha \rangle} {(E_{\alpha}-E_{\beta})^2}.
\end {align}
$|\alpha\rangle$ is the LL state with energy $E_{\alpha}$, S the area of the enlarged unit cell, $f_{\alpha,\beta}$ the Fermi-Dirac distribution functions, and $\mathbf{\dot{u}}_{x}$ the velocity operator along $\hat x$. The matrix elements of the velocity operators, which determine the available inter-LL transitions, are evaluated from the gradient approximation \cite{PRB50;17744}
\begin{align}
\langle \alpha  |\mathbf{\dot{u}}_{x}| \beta\rangle = \frac {1}{\hbar} \langle \alpha  |\frac {\partial H} {\partial k_x} | \beta\rangle \nonumber \\
\langle \alpha  |\mathbf{\dot{u}}_{y}| \beta\rangle = \frac {1}{\hbar} \langle \alpha  |\frac {\partial H} {\partial k_y} | \beta\rangle.
\end {align}

This approximation has been successfully used to investigate the optical spectra of carbon-related systems. The velocity matrix elements are dominated by the intralayer nearest-neighbor interaction,\cite{PCCP18;17597} so that the quantized mode of the initial state on the $A^l$ sublattice must be identical to that of the final state on the $B^l$ sublattice.\cite{PCCP17;26008} Apparently, the well-behaved, perturbed and undefined LLs of layered graphenes,  accompanied with various magnetic selection rules,\cite{CR4;7509} are expected to exhibit the rich and unique quantum conductivities.

\vskip 0.6 truecm
\par\noindent
\section{\bf Results and Discussion }
\vskip 0.1 truecm

\vskip 0.6 truecm
\par\noindent
\subsection{\bf Quantum Hall effect in sliding bilayer graphene }
\vskip 0.3 truecm

The essential electronic properties of sliding bilayer graphene are very sensitive to the changes in stacking configurations (Figs 2(a)-2(f)). AA bilayer graphene has two pairs of linear conduction and valence bands under the strong overlap (inset in Fig. 2(a)), while AB bilayer graphene possesses two pairs of parabolic bands with a weak band overlap near the Fermi energy (inset in Fig. 2(d)).\cite{ACSN;1465}
The dramatic transformation between the Dirac cones and the parabolic bands results in the strong hybridization of the two neighboring conduction (valence) bands.\cite{CR4;7509}
Particularly, the intermediate stackings of $\delta=b_0/8$ (inset in Fig. 2(b)), $\delta=6b_0/8$ (Fig. 2(c)), and $\delta=11b_0/8$ (Fig. 2(e)) present the special band structures, in which electronic states in the lower cone of the first pair and those of the upper cone of the second pair are strongly hybridized. An eye-shape stateless region is created near $E_F$ along $\hat{k}_x$ or $\hat{k}_y$.
The band structures are drastically changed even in the high-symmetry AA$'$ stacking ($\delta=1.5b_0$ in Fig. 2(f)). This system exhibits two pairs of titled cone structures with the non-vertical Dirac points.
The various electronic structures are magnetically quantized into the diverse LLs characterized by the rich and unique features. Two groups of valence and conduction LLs, the first and the second ones ($n_1^{c,v}$ and $n_2^{c,v}$; blue and red curves in Fig. 2), are initiated from the zero-field band-edge states. All the $B_0$-dependent LL energies exhibit the crossing behaviors of the well-behaved LLs.
On the other hand, the anticossing phenomenon will occur when two LLs possess certain identical components. The anticrossing energy spectra, arising from the perturbed and undefined LLs, are revealed in the lower-symmetry stacking systems, including the $\delta = b_0/8$, $\delta = 6b_0/8$, and $\delta = 11b_0/8$ ones.
Especially, for the latter two stacking configurations, there exist a lot of undefined LLs at higher energy range, clearly indicating the very frequent intergroup anticrossings.
In general, whether two or one groups of LLs make contributions to Hall conductivity depends on the range of energy overlap.\\

\begin{figure}[H]
\center
\rotatebox{0} {\includegraphics[width=0.9\linewidth]{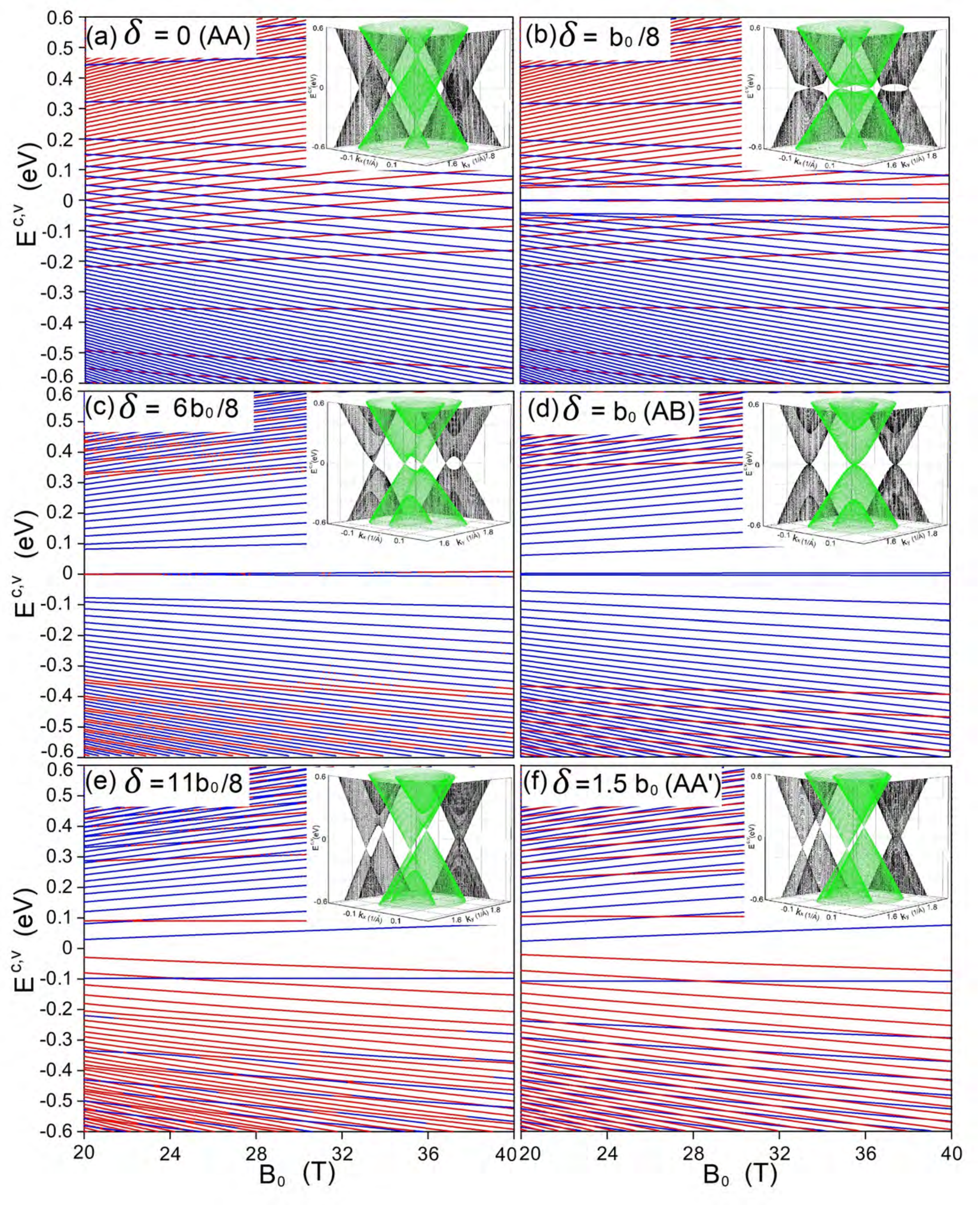}}
\label{Figure 2}
\end{figure}

\begin{figure}[H]
\center
\rotatebox{0} {\includegraphics[width=0.9\linewidth]{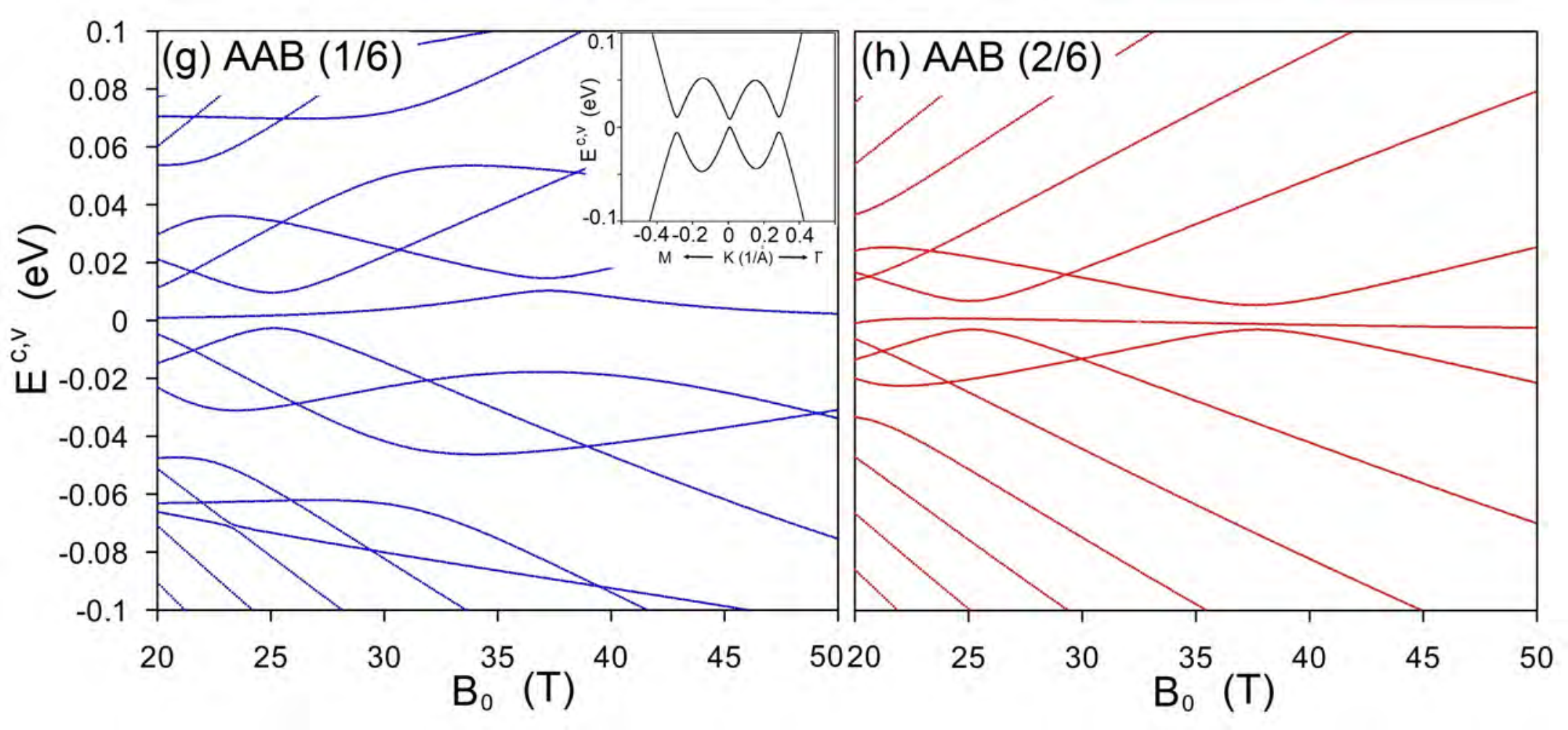}}
\caption{Magnetic-field-dependent Landau level energy spectra for sliding bilayer graphene with the shift of (a) $\delta=0$ (b) $\delta=b_{0}/8$, (c) $\delta=6b_{0}/8$, (d) $\delta=b_{0}$, (e) $\delta=11b_{0}/8$; (f) $\delta=1.5b_{0}$, and those of AAB-stacked trilayer graphene at (g) 1/6 and (h) 2/6 localization centers.}
\label{Figure 2}
\end{figure}

The QHE of AA bilayer graphene could be regarded as the superposition of those from two overlapping Dirac-cone structures; that is, only electronic transitions between two intragroup LLs exist and the available ones satisfy the selection rules of $\Delta n = \pm 1$. The $E_F$-dependent quantum Hall conductivity is quantized as $\sigma_{xy}=4me^2/h $ (Figs. 3(a)-(b)). The step structures present a unit height of $4e^2/h$, in which the plateaus are located at 0, $\pm 4e^2/h$, $\pm 8e^2/h$ and so on.
A wide plateau, with the insulating behavior, covers zero energy (Fig. 3(a) at $B_0=40$ T), being related to the competitive relation between two groups of LLs. In other words, the Hall conductivity is vanishing at the neutrality point, similar to that of monolayer graphene. In monolayer system, the available interband transitions of the $\Delta n = n^c -n^v=1$ and $-1$ selection rules, respectively, possess the positive and negative Hall conductivities. Moreover, the valence and conduction LLs are symmetric about $E_F=0$, leading to the vanishing conductivity at zero energy. As for AA bilayer graphene, the total conductivity is the combination of electronic transitions in two separated LL groups which are almost symmetric about $E_F=0$. At the neutrality point, the available transitions of the first group (blue lines) are dominated by the valence LLs while those of the second group (red lines) are governed by the conduction LLs. The former and the latter, respectively, satisfy the selection rules of $\Delta n_1^{v\rightarrow v} =  -1$ and $\Delta n_2^{c\rightarrow c} =  1$; therefore, they cancel each other out.
In general, a plateau of 4e$^2$/h height appears when a certain LL becomes occupied or unoccupied during the variation of $E_F$ (Figs. 3(a) and 3(b)). This might be changed into a double step for the merged two LLs in distinct groups (the intergroup LL crossing), e.g., $8e^2/h$ at $E_F\sim 0.026$ eV (Fig. 3(a)).
It is worth noting that the quantized LLs at two Dirac points in AA bilayer system ($n_1=0$ and $n_2=0$) are fully occupied/unoccupied, being different from the half-occupied n = 0 LL in monolayer system. Apparently, the quantum Hall conductivity does not present the steps of $2e^2/h$ height near ${E_F=0}$ as in monolayer graphene.\cite{N438;197, N438;201}\\

\begin{figure}[H]
\center
\rotatebox{0} {\includegraphics[width=0.9\linewidth]{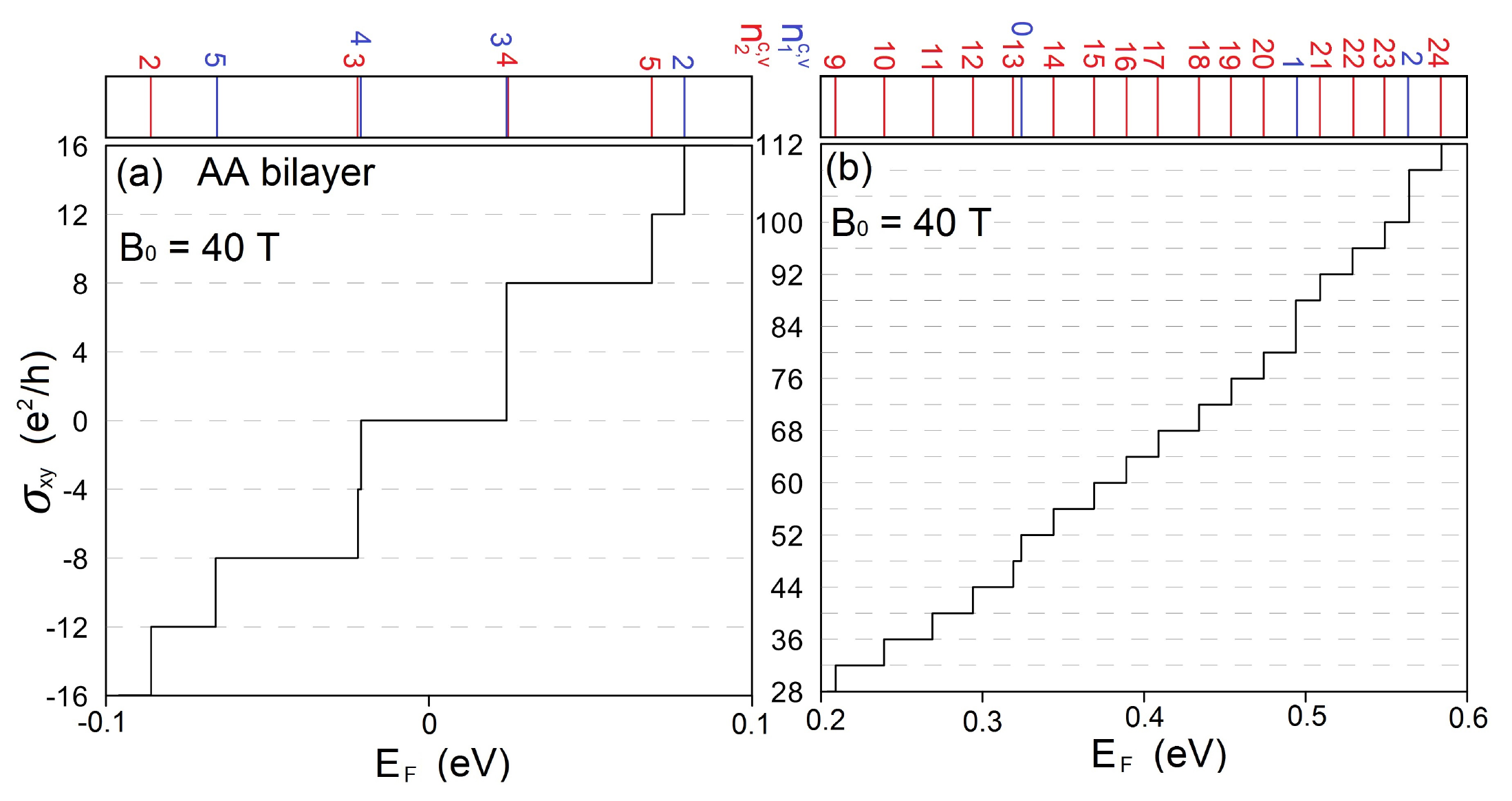}}
\caption{Fermi-energy-dependent Hall conductivities of AA bilayer graphene at (a) lower- and (b) higher-energy ranges for $B_0$ = 40 T.}
\label{Figure 3}
\end{figure}

The quantum Hall conductivity in AB bilayer graphene is quantized as $\sigma_{xy}=4m e^2/h$ or $\sigma_{xy}=4m^{'} e^2/h$ (Figs. 4(a) and 4(b)), depending on the strength of magnetic field.
The low $E_F$-dependent quantum Hall conductivity exhibits the usual step structures, in which the plateaus are higher while their widths are declined with the growth of energy.
Especially, there appears a very narrow plateau near zero energy (Fig. 4(a)). Such plateau is observable at the sufficiently large field strength.
It is induced by a small energy spacing between the $n_1=0$ and $n_1=1$ LLs of the first group. This energy spacing is monotonically dependent on the field strength, as shown in Fig. 2(d). It will be infinitesimal and even disappears below the critical field strength ($B_0 < 20$ T, e.g., 12T in the inset), leading to a double step in the Hall conductivity. These results are in agreement with the previous predictions and measurements.\cite{NP2;177, PRB78;033403, PRL96;086805}
At higher energy range ($E_F > 0.35$ eV; Fig. 4(b)), the quantum conductivity is dominated by two groups of LLs (blue and red lines). There are four categories of electronic transitions, including the intragroup ($n_1^{c} \rightarrow n_1^{c}$, $n_2^{c} \rightarrow n_2^{c}$) and intergroup ones ($n_1^{c} \rightarrow n_2^{c}$, $n_2^{c} \rightarrow n_1^{c}$).\cite{ACSN;1465} The number of steps is significantly enhanced as shown in Fig. 4(b); furthermore, the change in plateau widths no longer follows a simple relation with $E_F$. Specifically, the available intergroup transitions, with the same quantized  mode on the $A^l$ ($B^l$) sublattice (Eq. (3)), only  happen in  two LLs with a large energy spacing. They only make negligible contributions, since the quantum Hall conductivity is inversely proportional to the square of energy difference between the initial and final states (Eq. (2)). This is in great contrast with the frequency-dependent optical excitation spectra.\cite{ACSN;1465} That is to say, the QHE in AB bilayer graphene can be considered as the superposition of the two intragroup transition channels. \\

\begin{figure}[H]
\center
\rotatebox{0} {\includegraphics[width=0.9\linewidth]{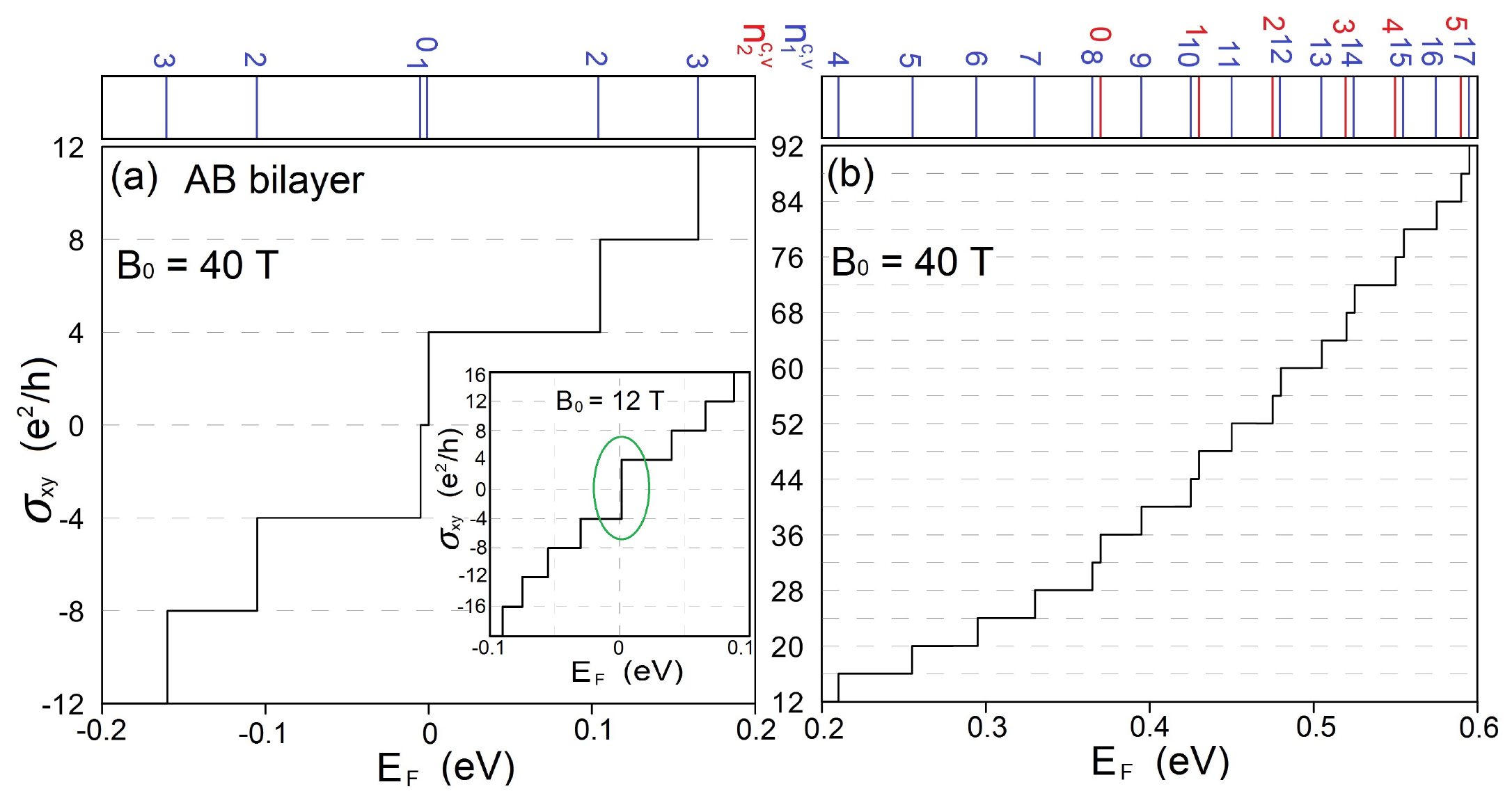}}
\caption{Fermi-energy-dependent Hall conductivities of AB bilayer graphene at (a) lower- and (b) higher-energy ranges for $B_0$ = 40 T.}
\label{Figure 4}
\end{figure}

The $B_0$-dependent Hall conductivity in AA bilayer graphene exhibit the wells, the monolayer-like or non-monotonous staircases, and the composite ones (Figs. 5(a)-5(d)), depending on the $B_0$-dependent LL energy spectrum (Fig. 2(a)). In particular, there are four types of step structures corresponding to the different LL spectral regions. At low-energy range ($|E_F| <$ 0.1 eV), the $n_1^{v}$ valence LLs (blue curves in the inset of Fig. 5(a)) and the $n_2^{c}$ conduction LLs (red curves) cross one another and thus create the rhombus pattern area, being accompanied with the well-like quantum conductivity (Fig. 5(a)). In fact, the electrical conductivity only possesses two values of 0 and $4e^2/h$. Since the intragroup transitions of the $n_1^{v}$ LLs and those of the $n_2^{c}$ ones present the opposite conductivities ($+$ $\&$ $-$), the Hall conductivity is invariant at the overlap point of two LLs from distinct groups during the variation of field strength. At $E_F$ = 0, the number of the unoccupied $n_1^{v}$ LLs is identical to that of the occupied $n_2^c$ ones, resulting in zero conductivity. When $E_F$ is slightly increased ($E_F=0.01$ eV in Fig. 5(a)), the variation of $B_0$ can create the asymmetric LL distribution. This is responsible for the steps of $4e^2/h$. For example, according to the range of 22.5 T $\le B_0 \le 24$ T, the $n_1^{v} = \color{blue}{6}$ LL is changed into the occupied state at the initial $B_0$, while  the $n_2^{c} = \color{red}{6}$ LL become the unoccupied one at the final $B_0$ (inset). These two LLs, respectively, create and destroy a step of $4e^2/h$; that is, they make opposite contributions to quantum conductivity.
With the increase of the Fermi energy, the well-like and staircase steps could co-exist in the range of 0.1 eV $\leq E_F <$ 0.2 eV.
The $n_1^v$ and $n_2^c$ LLs have the highly non-equivalent distributions. For example, at $E_F=0.1$ eV (Fig. 5(b)), the unoccupied $n_2^c=\color{red}{11}$, occupied $n_1^v=\color{blue}{3}$  and unoccupied $n_2^c=\color{red}{10}$ LLs are, respectively, revealed at $B_0$=21, 22.5; 23.5 T's, so that the well-like structure occurs there. However, only the unoccupied $n_2^c$ LLs appear in the range of 23.5 T $\le B_0 \le$ 33.5 T, indicating the further decrease of quantum conductivity and thus the formation of staircase.
Specifically, as for the Fermi energy above the $n_1^{v} = \color{blue}{0}$ LL ($E_F=0.35$ eV in Fig. 5(c)), only the $n_2^{c}$ LLs exhibit the dramatic transformation in the increase of field strength (inset). As a result, the $B_0$-dependent quantum conductivity behaves similarly to that of monolayer graphene.\cite{PRL95;146801} The plateaus are decreased by a step of $4e^2/h$ and their widths are gradually increased whenever an extra $n_2^c$ LL becomes unoccupied. The enhancement of the plateau width directly reflects the $B_0$-difference between two neighboring unoccupied $n_2^c$ LLs.
At the sufficiently high energy ($E_F> 0.4$ eV), the $n_1^{c}$ and $n_2^{c}$ conduction LLs overlap together. These two groups of LLs possess the similar dependencies on the magnetic field; therefore, the number of the step structures is greatly enhanced ($E_F=0.55$ eV in Fig. 5(d)). Particularly, with the growth of $B_0$ within the range of 20 T $\rightarrow$ 40 T, the Hall conductivity shows more step structures as the $n_1^{c}$ LLs come to exist (green and red arrows in Fig. 5(d)). A very narrow plateau width near 25 T (green arrow) corresponds to the field-strength difference in the unoccupied $n_1^{c} = \color{blue}{3}$ and $n_2^{c} = \color{red}{38}$ LLs (green circle). Especially, a plateau of $132e^2/h$ is absent at the crossing point of the unoccupied $n_1^{c}=\color{blue}{2}$ and $n_2^{c} = \color{red}{25}$ LLs (inset), leading to a step of height $8e^2/h$ (red arrow).

\begin{figure}[H]
\center
\rotatebox{0} {\includegraphics[width=0.9\linewidth]{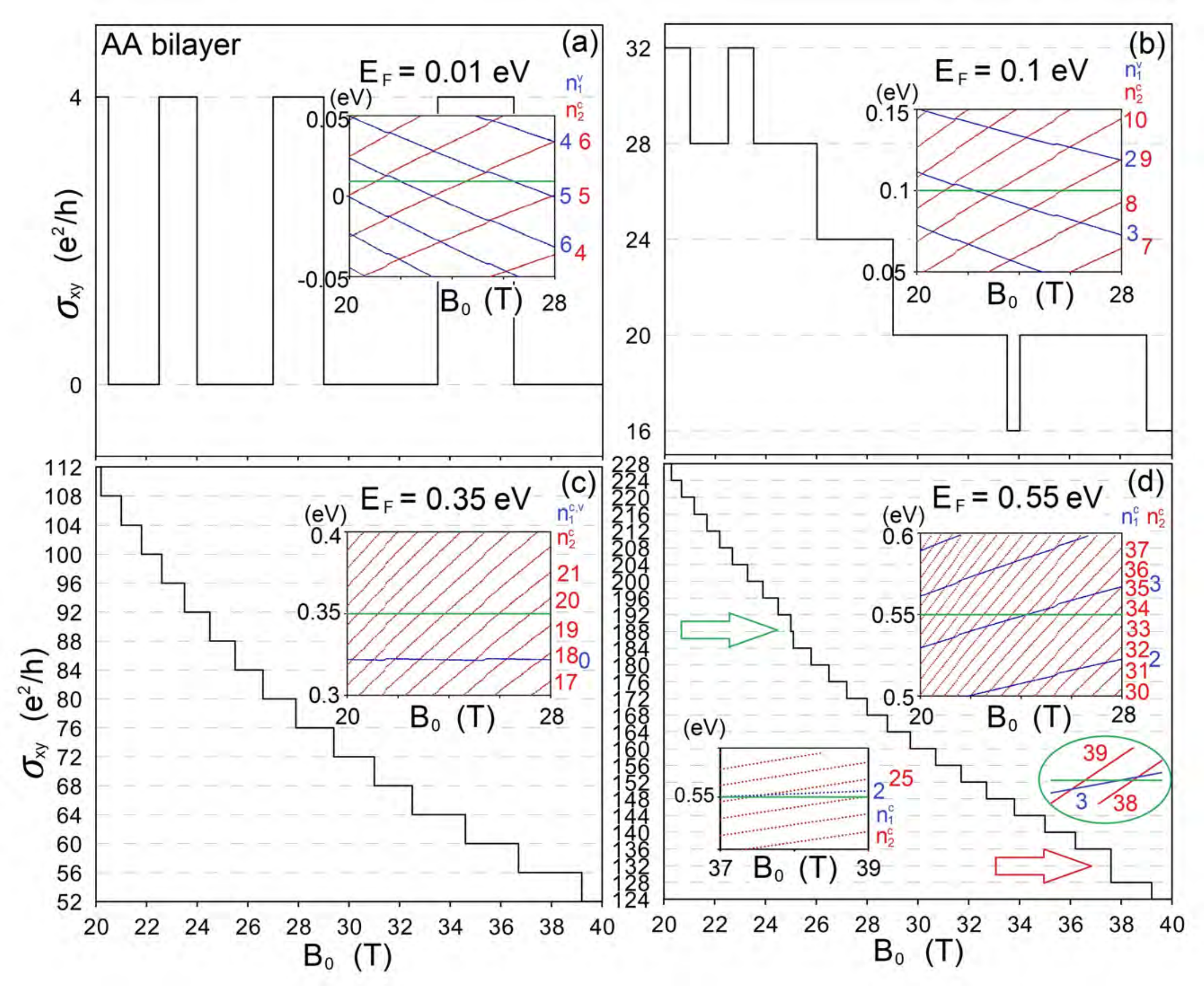}}
\caption{Magnetic-field-dependent Hall conductivities of AA bilayer graphene at (a) $E_F$ = 0.01 eV, (b) $E_F$ = 0.1 eV, (c) $E_F$ = 0.35 eV and (d) ${E_F=0.55}$ eV. Also shown in the insets are the $B_0$-dependent LL energy spectra, with the Fermi level indicated by the green line.}
\label{Figure 5}
\end{figure}

The Hall conductivity of AB bilayer graphene is monotonically decreased for the rise of field strength, as shown in Figs. 6(a) and 6(b).
At the lower Fermi energy, the step structures, as shown in Fig. 6(a) for $E_F=0.2$ eV, are dominated by the electronic transitions between the low-lying LLs of the first group (blue curves in the inset).\cite{ACSN;1465} More $n_1^c$ LLs become unoccupied, leading to the monolayer-like QHE.
However, at higher energy range ($E_F > 0.35$ eV), the overlaps of the two LL groups remarkably enrich the Hall conductivity spectrum, e.g., $E_F$=0.5 eV in Fig. 6(b). Although the step structures are the combination of all the available electronic transitions among two groups of LLs, only the intragroup ones are dominant (as discussed earlier). Consequently, a simple relation between quantum conductivity and field strength is absent. The small energy spacings between the $n_1^{c}$ and $n_2^{c}$ LLs induce a number of narrow plateaus as indicated by the blue arrows. Furthermore, a double step of $8e^2/h$ occurs at around 33.7 T (red arrow), owing to the overlap of the $n_1^{c} = \color{blue}{15}$ and $n_2^{c}=\color{red}{3}$ LLs (inset in Fig. 6(b)). As can be predicted from the crossing behavior of the $B_0$-dependent LL spectrum (Fig. 2(d)), the weaker the field strength, the more frequently will the double-height steps appear.\\

\begin{figure}[H]
\center
\rotatebox{0} {\includegraphics[width=0.9\linewidth]{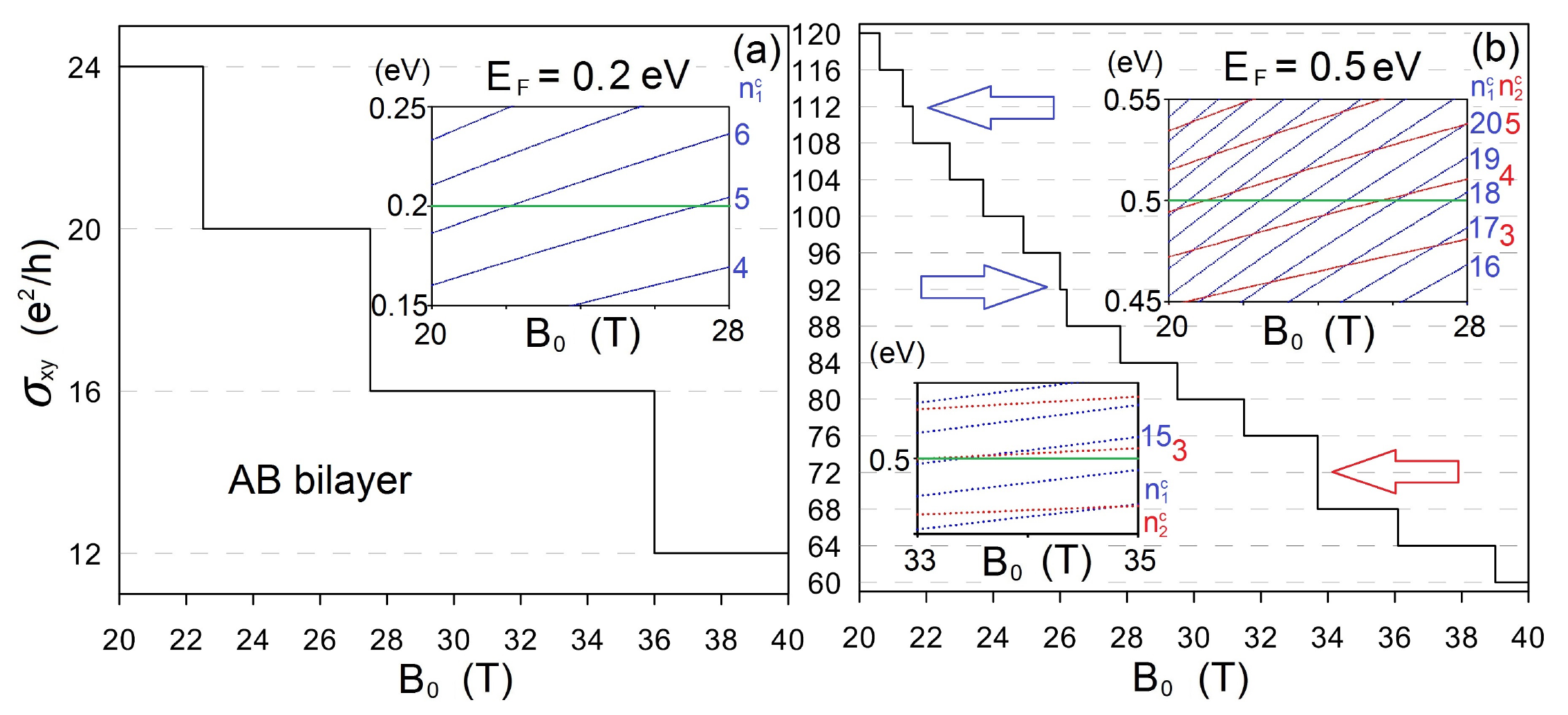}}
\caption{Magnetic-field-dependent Hall conductivities of AB bilayer graphene at (a) $E_F$ = 0.2 eV and (b) $E_F$ = 0.5 eV. Insets show the occupied or unoccupied LLs.}
\label{Figure 6}
\end{figure}

The quantum conductivities are diversified by the shift-dependent stacking configurations. The lower-symmetry bilayer system of $\delta=b_0/8$ shows the unusual step structure, as clearly indicated in Fig. 7(a).
The low-lying plateaus are quantized as a special sequence of (-9, -4, -3, 4, 7), leading to the exclusive step heights of $5e^2/h$, $e^2/h$, $7e^2/h$, and $3e^2/h$. Such odd-integer steps are associated with the LLs of $n_2^{c}=\color{red}{2}$, $\color{red}{3}$, $\color{red}{4}$, and $\color{red}{5}$ (red lines) and $n_1^{v}= \color{blue}{2}$, $\color{blue}{3}$, $\color{blue}{4}$ and $\color{blue}{5}$ (blue lines). These LLs are quantized from the hybridized electronic states near the eye-shape region of the band structure (inset in Fig. 2(b)). There exist extra inter-group electronic transitions among these LLs which satisfy the selection rules of $\Delta n = 0$ $\& \pm 2$, e.g., $n_2^{c}={\color{red}{3}} \rightarrow n_1^{v}={\color{blue}{3}}$, $n_2^{c}={\color{red}{4}} \rightarrow n_1^{v}={\color{blue}{2}}$, and others. Especially, only the transitions associated with $\Delta n = 2$ rule will contribute a positive quantum conductivity.
As a result, the low-energy QHE is mainly determined by the intra- and inter-group LL transition channels.
When $E_F$ crosses zero from a negative value, the $n_2^{c}=\color{red}{3}$ becomes occupied. The electronic transition from this LL to the $n_1^{v}={\color{blue}{3}}$ is available, and it generates the quantum conductivity of $-3e^2/h$. The zero-energy Hall conductivity originates from two intragroup transition channels of ($n_2^{c}={\color{red}{3}} \rightarrow n_2^{c}={\color{red}{4}}$ $\&$ $n_1^{v}={\color{blue}{4}} \rightarrow n_1^{v}={\color{blue}{3}}$) and the $n_2^{c}={\color{red}{3}} \rightarrow n_1^{v}={\color{blue}{3}}$ intergroup transition. The former cancel each other out, being responsible for the $-3e^2/h$ conductivity (red arrow). This system is thoroughly different from AA and AB bilayer graphenes with zero conductivities at the neutrality point.
With the Fermi energy above the $n_1^{v}={\color{blue}{3}}$ LL, the intragroup transitions of ($n_1^{v}={\color{blue}{3}} \rightarrow n_1^{v}={\color{blue}{2}}$ and $n_2^{c}={\color{red}{3}} \rightarrow n_2^{c}={\color{red}{4}}$) make the main contribution, and they create the quantum conductivity of $4e^2/h$ (blue arrow). These two special plateaus ($-3e^2/h$ $\&$ $4e^2/h$) result in a step height of $7e^2/h$.
The competitive or cooperative relation between the intragroup and intergroup transitions (two intragroup ones) is also revealed at lower and higher Fermi energies.
The dominating transitions of ($n_1^{v}={\color{blue}{5}} \rightarrow n_1^{v}={\color{blue}{4}}$, $n_2^{c}={\color{red}{2}} \rightarrow n_1^{v}={\color{blue}{4}}$, $n_2^{c}={\color{red}{2}} \rightarrow n_2^{c}={\color{red}{3}}$), ($n_1^{v}={\color{blue}{4}} \rightarrow n_1^{v}={\color{blue}{3}}$, $n_2^{c}={\color{red}{2}} \rightarrow n_2^{c}={\color{red}{3}}$, $n_1^{v}={\color{blue}{4}} \rightarrow n_2^{c}={\color{red}{4}}$) and ($n_1^{v}={\color{blue}{3}} \rightarrow n_1^{v}={\color{blue}{2}}$, $n_2^{c}={\color{red}{4}} \rightarrow n_2^{c}={\color{red}{5}}$, $n_2^{c}={\color{red}{4}} \rightarrow n_1^{v}={\color{blue}{2}}$), respectively, induce the ${-9e^2/h}$, ${-4e^2/h}$ and ${7e^2/h}$ quantum conductivities at ${E_F=-0.046}$, 0 and 0.06 eV's
(purple, black and green arrows).

 \begin{figure}[H]
\center
\rotatebox{0} {\includegraphics[width=0.9\linewidth]{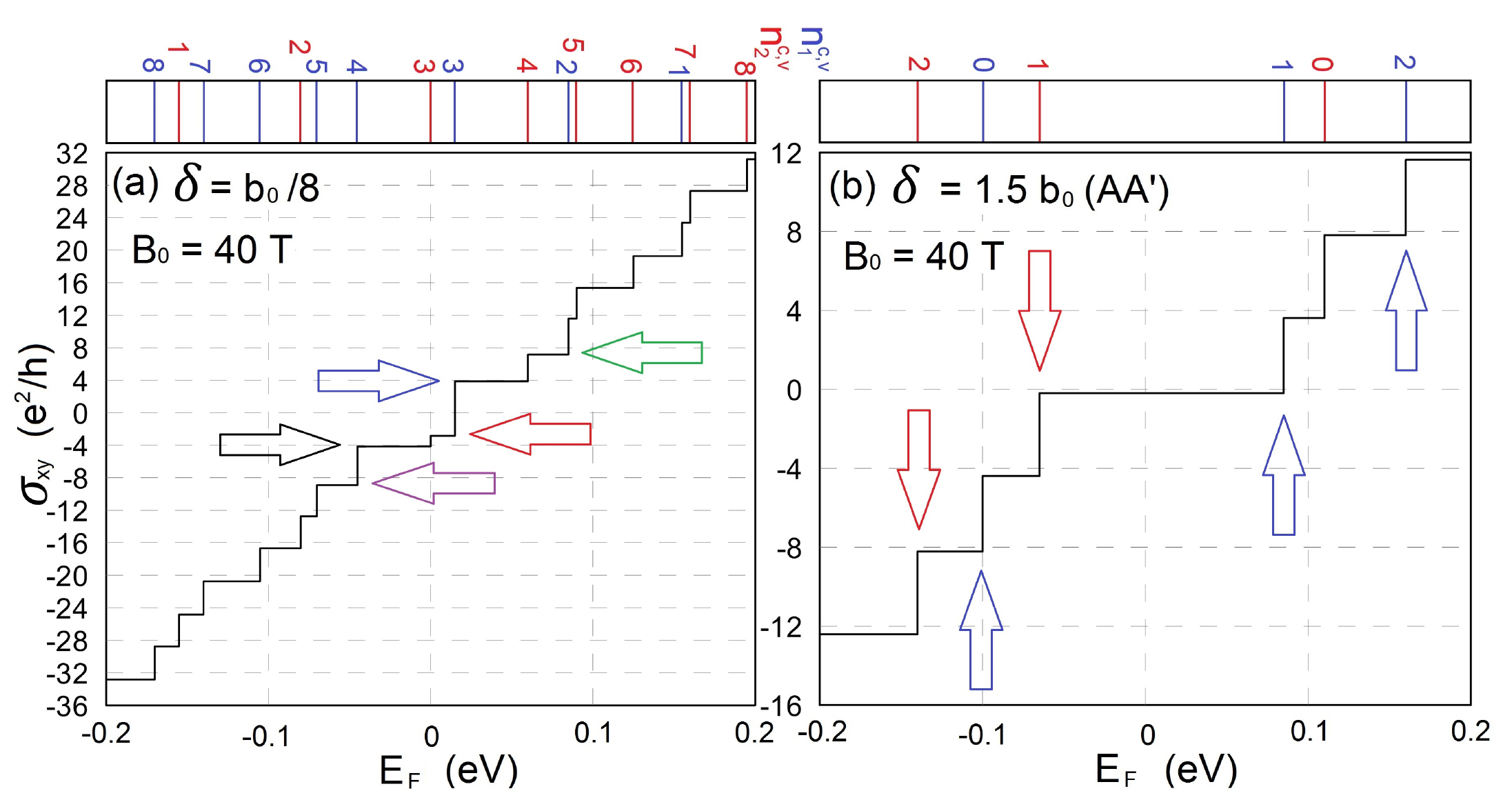}}
\caption{Fermi-energy-dependent Hall conductivity of sliding bilayer graphene with the shift of (a) $\delta=b_0/8 $ and (b) $\delta= 1.5 b_0 $ at $B_0$ = 40 T.}
\label{Figure 7}
\end{figure}

Specifically, the high-symmetry AA$^\prime$ stacking presents the non-integer plateaus (Fig. 7(b)), being in great contrast to AA bilayer (Fig. 3(a)). Two groups of well-behaved LLs, which are quantized from the tilted Dirac cones (Fig. 2(f)), only contribute the intragroup LL transitions of $\Delta n=\pm 1$ to quantum conductivity.
When an extra $n_1^{c,v}$ LL becomes occupied in the increase of $E_F$, the quantum conductivity (blue arrows in Fig. 7(b)) is enhanced by a height of $3.8e^2/h$. On the other hand, the occupation of one $n_2^{c,v}$ LL can create a $4.2e^2/h$-height step (red arrows).
The non-integer QHE might be associated with the tilted Dirac cones, in which the localization centers of LLs gradually change with the increasing energy.\cite{CR4;7509}
Also, the non-fixed centers could induce a small conductivity of ${0.2e^2/h}$ near the neutrality point.\\

As for the undefined LLs in $\delta =6b_0/8$ and $11b_0/8$ bilayer stackings (Figs. 8(a) and 8(b)), their quantum conductivities are worthy of a detailed examination. The former presents the $n_1^{c,v}$ conduction and valence LLs, with the perturbed modes, in the low-lying spectral region under the absence of the $n_2^{c,v}$ LLs (Fig. 2(c)).
At higher/deeper energy range where the $n_2^{c,v}$ LLs come to exist, two groups of LLs are continuously anticrossed together, as shown in the inset of Fig. 8(a). The localization modes of each LL cannot be well defined, since they vary with the field strength (Fig. 2(c)). Such highly degenerate electronic states without a specific main mode are classified as the undefined LLs.\cite{CR4;7509}
It is noted that $n_2^{c,v}$ in Fig. 8(a) (also in Fig. 8(b)) only denotes the ordering of LLs, but not the quantum numbers. The available transition channels cover the intragroup and intergroup ones in the absence of selection rule, while they are dominated by the neighboring LLs.
The quantum Hall conductivity is monotonically declined with the increase of $B_0$, similar to that of AB bilayer graphene (Figs 6(a)-6(b)). Since the undefined LLs do not have the crossing behavior, all the plateaus are quantized as single steps of height $4 e^2/h$ without the double structures. When one $n_1^c$/$n_2^c$ LL is getting unoccupied, the related transition channels can reduce an integer quantum conductivity of $4e^2/h$.
Moreover, an inverse relation between plateau width and $B_0$ is absent because of the frequently irregular anticrossings.
The similar quantum conductivity is revealed by the undefined LLs of the $\delta=11b_0/8$ stacking. It might have some narrow plateaus (black arrows in Fig. 8(b)), e.g., the plateau at $B_0 \sim$ 28 T arising from the $n_1^{c}=\color{blue}{13}$ and $n_2^{c} = \color{red}{3}$ LLs.
Such plateaus directly reflect the very weak anticrossings.

\begin{figure}[H]
\center
\rotatebox{0} {\includegraphics[width=1.0\linewidth]{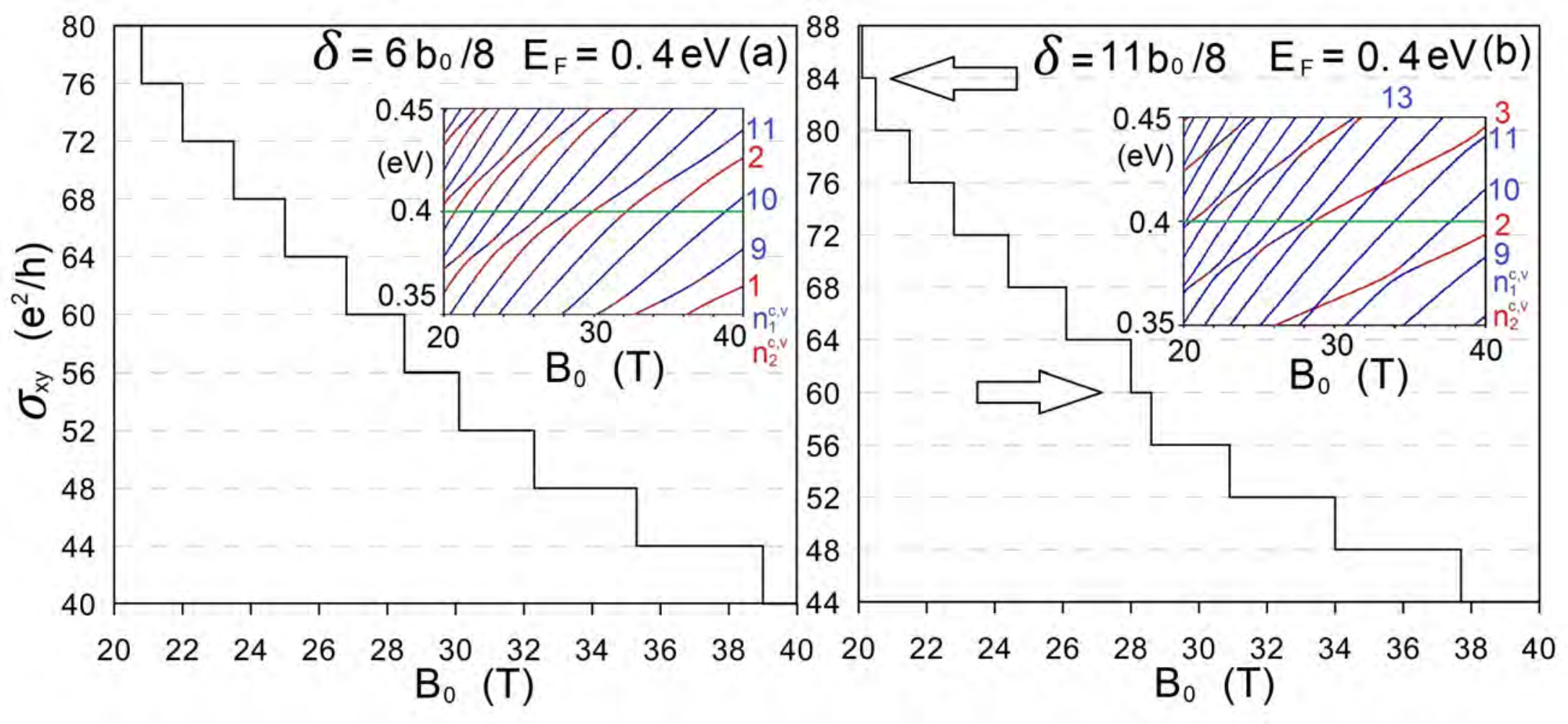}}
\caption{Magnetic-field-dependent Hall conductivities of (a) $\delta=6b_0/8$ and (b) $\delta=11 b_0/8$ stacking systems. Also shown in the insets are those of the LL energy spectra.}
\label{Figure 8}
\end{figure}

\vskip 0.6 truecm
\par\noindent
\subsection{\bf Quantum Hall effect in trilayer graphene}
\vskip 0.3 truecm

In AAB-stacked trilayer graphene, the oscillatory energy bands of the first pair (inset in Fig. 2(g)) are magnetically quantized into the unique LLs. Especially, the low-lying LLs, with perturbed modes, present the frequently crossing and anticrossing behaviors, as clearly shown in Figs. 2(g) and 2(h).
All the LLs are doubly degenerate, since the destruction of the $z$-plane mirror symmetry leads to the non-equivalence of two valleys (or two distinguishable localization centers of 1/6 and 2/6 in an enlarged unit cell; details in Ref. 40).
Apparently, the feature-rich LLs will induce the unusual QHE.
The $E_F$-dependent quantum Hall conductivity exhibits the step height of $2e^2/h$ instead of the usual one of $4e^2/h$  and the irregular widths of  plateaus (Fig. 9(a)).
At low energy, there also exist certain double-height steps of $4e^2/h$.
Those steps are closely related to the crossings of entangled LLs around the neutrality point (Figs. 2(g) and 2(h)).
As a result of the splitting energy spectra, the LLs localized at 1/6 and 2/6 centers are very different in the quantum conductivities. As $E_F$ approaches zero energy, a plateau appears at $B_0$ = 40 T for the 1/6 center (blue curve in Fig. 9(b)), but two neighboring steps correspond to the 2/6 center (red curve).
The absence/presence of the step structure depends on whether the anti-crossing LLs could survive there. For example, at the 1/6 center, the anti-crossing $n_1^{c,v}=0$ and $n_1^{c}=3$ LLs away from $E_F=0$ (inset in Fig. 10(a)) accounts for the absence of zero-energy step.
On the other hand, at the 2/6 center, the $n_1^{c,v}=0$ and $n_1^v=1$ LLs continuously become occupied at $E_F=-3$ meV and 0, so that the dominant  transition channels of
${n_1^{v}=1\rightarrow\,n_1^{c,v}=0}$ and ${n_1^{c,v}=0\rightarrow\,n_1^c=1}$ create the neighboring narrow-width steps.
Concerning the neutrality point, the competition between these two localization centers lead to zero conductivity.
In short, the splitting and anti-crossing energy spectra have generated the complex quantum structures  with the half-reduced conductivities (Fig. 9(a)).

\begin{figure}[H]
\center
\rotatebox{0} {\includegraphics[width=0.9\linewidth]{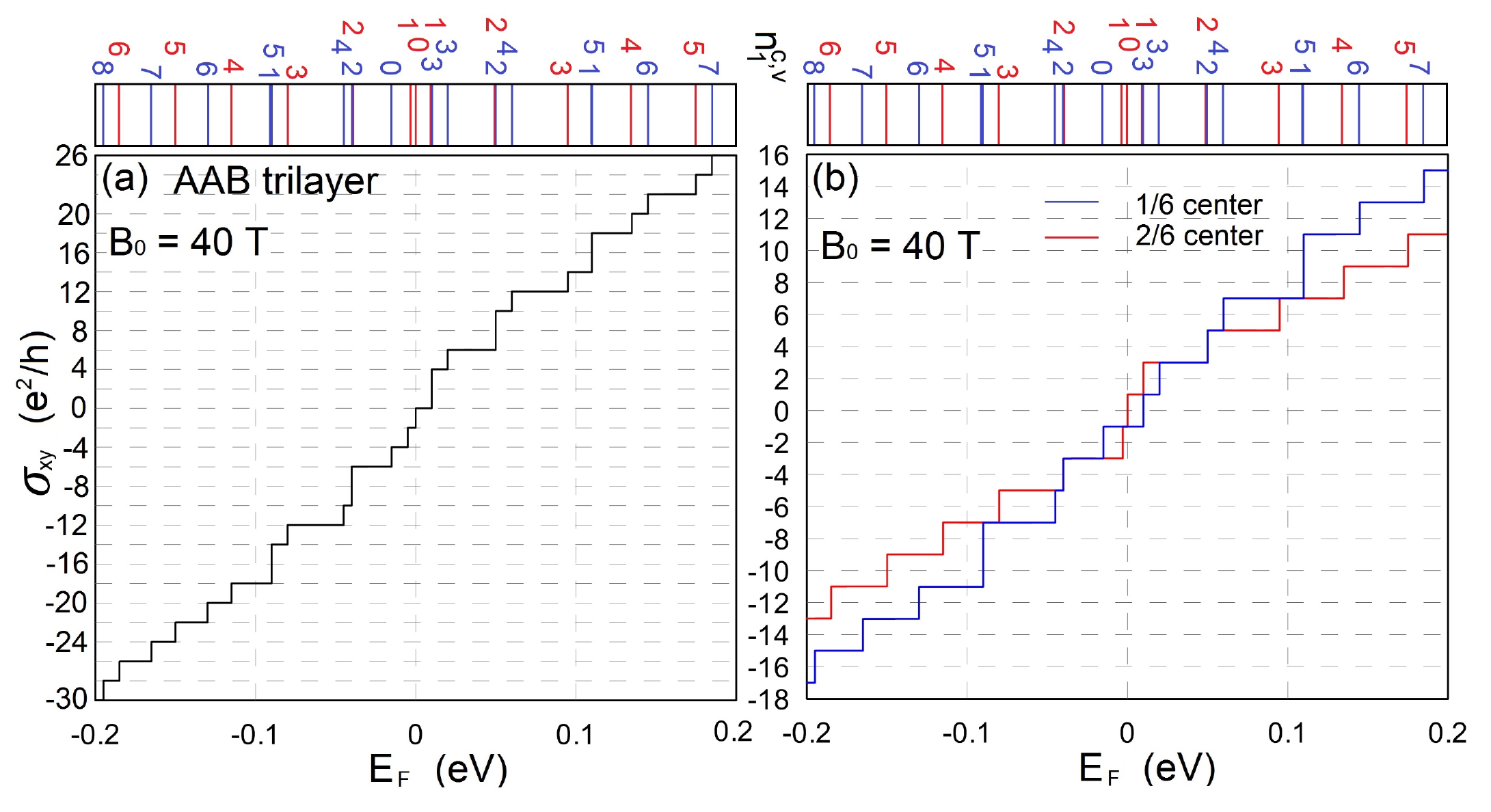}}
\caption{Fermi-energy-dependent Hall conductivity of (a) AAB-stacked trilayer graphene, being related to (b) two distinct localization centers.}
\label{Figure 9}
\end{figure}

The $B_0$-dependent Hall conductivity, being associated with the low-lying LLs, present the extraordinary quantum structures (Fig. 10(c)). The abnormal plateaus are absent in the sliding bilayer system (Figs. 5, 6 $\&$ 8). They mainly originate from the intragroup anticrossing LLs (or the oscillatory energy bands). For the 1/6 center, the positive/negative quantum conductivity appears and then disappears during a certain range of $B_0$, as shown in Fig. 10(a). When the magnetic field reaches the critical value of ${B_0=24.75}$ T, the ${n_1^c=4}$ LL becomes occupied and the dominating ${n_1^{v}=0\rightarrow\,n_1^{c}=4}$ transition is thus forbidden (inset).
The original channel makes an contribution of ${-2e^2/h}$, so an upward step appears there. And then, a higher plateau is changed into the original one at another critical field of ${B_0=25.5}$ T. With a further increase of field strength, the opposite quantum structure comes to exist in the range of 37 T${\le\,B_0\le\,38.5}$ T. The reduced and enhanced quantum conductivities, respectively, correspond to the disappearance and recovery of the  ${n_1^{v}=0\rightarrow\,n_1^{c}=3}$ channel.
It should be noticed that the extra selection rules of ${\Delta\,n=\pm\,3}$ and ${\pm\,4}$ belong to the available transition channels because of the co-existent main and side modes in the perturbed LLs. On the other hand, the LLs localized at ${2/6}$ center only present the well-like quantum structures (Fig. 10(b)), i.e., the same enhancement and reduce of quantum conductivity occurs continuously. Each structure is dominated by the same LLs within two neighboring critical fields (the ${n_1^{v}=0\rightarrow\,n_1^{c}=2}$ or ${n_1^{v}=0\rightarrow\,n_1^{c}=1}$ transition in the inset).\\

\begin{figure}[H]
\center
\rotatebox{0} {\includegraphics[width=0.9\linewidth]{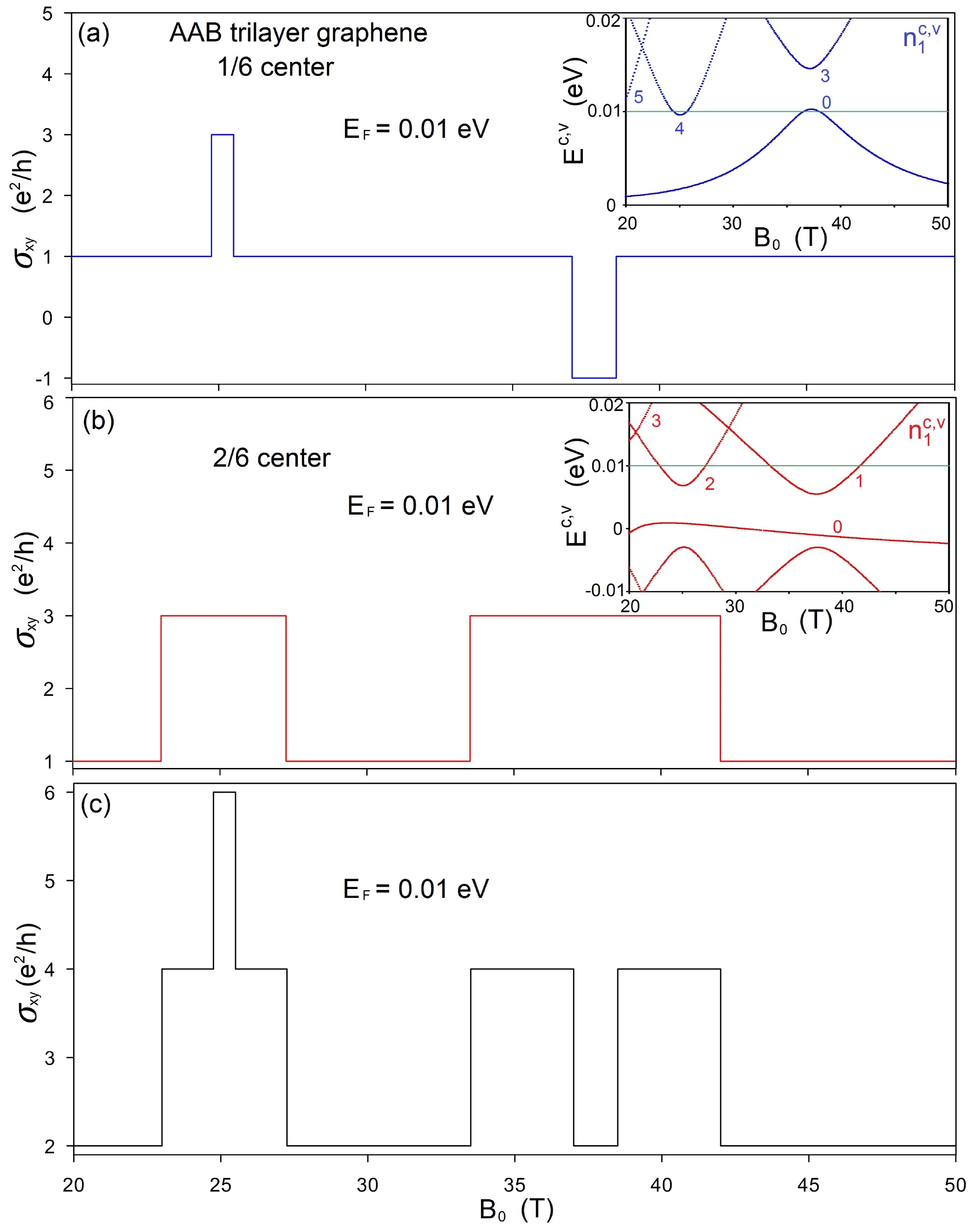}}
\caption{Magnetic-field-dependent Hall conductivities of AAB-stacked trilayer graphene at (a) 1/6 and (b) 2/6 localization centers. The total conductivity is shown in (c).}
\label{Figure 10}
\end{figure}

There exist certain important differences in quantum conductivities among the AAB-, ABA- and ABC-stacked trilayer graphenes. The step height of ${2e^2/h}$ in a wide energy range, the localization-dependent plateau structures, and the irregular $B_0$-dependence are absent in the latter two systems, since they have the higher-symmetry stacking configurations. In general, ABA and ABC have the monolayer-like steps with ${4e^2/h}$ height except that the unusual
quantum conductivities appear at a narrow energy range covering the neutrality point.\cite{N7;621, NP7;948, NP7;953} Such differences clearly illustrate the stacking-diversified QHE; that is, the stacking-dependent atomic interactions dominates the magnetic quantization and thus the magneto-transport properties.\\

\vskip 0.6 truecm
\par\noindent
\section{\bf Concluding Remarks}
\vskip 0.3 truecm

We have used the generalized tight-binding model and the Kubo formula to investigate the unusual QHE in high- and low-symmetry few-layer graphenes, especially for the identification of the selection rules from the former. This method could be further developed to explore the emergent 2D materials, e.g., silicene, germanene, tinene, phosphorene, and Mo$S_2$. The Fermi-energy- and magnetic-field-dependent quantum Hall conductivities are greatly diversified by various stacking configurations, since they might arise from the intragroup and intergroup LL transitions with/without the monolayer-like and extra selection rules. Three kinds of LLs and the crossing, anti-crossing and spitting  energy spectra are responsible for the diverse transport properties, such as, the integer and non-integer conductivities, the splitting-created reduction and complexity of quantum conductivity, the vanishing and non-zero conductivities at the neutrality point, the well-like, staircase and composite plateau structures, the distinct step heights, and the abnormal field dependence. These could be examined by magneto-transport measurements.\cite{N438;197, N438;201, PRB78;033403, NP2;177, N7;621, NP7;948}\\

As to the static inter-LL transitions, the bilayer AA, AB and AA$^\prime$ stackings present two kinds of intragroup transition channels under the normal selection rule of $\Delta n=\pm 1$, while the extra intergroup ones and selection rules might be revealed in the other low-symmetry systems.
The non-integer quantum conductivities only exists in the AA$^\prime$ stacking with the non-fixed LL localization center, covering the step heights of ${3.8e^2/h}$ and ${4.2e^2/h}$. Most of systems exhibit the monolayer-like step height of ${4e^2/h}$ even for the undefined LLs, such as AA, AB, ${\delta\,=6b_0/8}$ and ${\delta\,=11b_0/8}$ stackings. Moreover, the exclusive step heights of ${e^2/h}$, ${3e^2/h}$, ${5e^2/h}$ $\&$ ${7e^2/h}$ are observed in the ${\delta\,=1/8}$ stacking, in which a non-negligible conductivity appears near the neutrality point. Only the AA stacking has four types of $B_0$-dependent plateau structures (wells, monolayer-like or non-monotonous staircases, and composite ones), mainly owing to  the strong overlap of two Dirac-cone structures (LL energy spectra).
Specifically, the  AAB-stacked trilayer graphene exhibits the complex plateau structures with ${2e^2/h}$ height within a wide energy range and the irregular $B_0$-dependence, directly reflecting the splitting and  anti-crossing and LL energy spectra and the perturbed localization modes. These features are absent in ABA- and ABC-stacked  trilayer systems.\cite{N7;621, NP7;948, NP7;953} \\

\par\noindent {\bf Acknowledgments}

This work was supported in part by the Ministry of Science and Technology of Taiwan,
the Republic of China, under Grant No. MOST 105-2112-M-006 -002 -MY3.

\newpage

\renewcommand{\baselinestretch}{0.2}

\end{document}